# A Personalized Method for Calorie Consumption Assessment

## Yunshi Liu

College of Information Science and Engineering, Ritsumeikan University, Shiga, Japan
is0388ik@ed.ritsumei.ac.jp

## Pujana Paliyawan and Takahiro Kusano

Graduate School of Information Science and Engineering, Ritsumeikan University, Shiga, Japan
pujana.p@gmail.com, is0212kf@ed.ritsumei.ac.jp

## Tomohiro Harada and Ruck Thawonmas

College of Information Science and Engineering, Ritsumeikan University, Shiga, Japan
{harada@ci, ruck@is}.ritsumei.ac.jp

**Abstract**

This paper proposes an image-processing-based method for personalization of calorie consumption assessment during exercising. An experiment is carried out where several actions are required in an exercise called broadcast gymnastics, especially popular in Japan and China. We use Kinect, which captures body actions by separating the body into joints and segments that contain them, to monitor body movements to test the velocity of each body joint and capture the subject's image for calculating the mass of each body joint that differs for each subject. By a kinetic energy formula, we obtain the kinetic energy of each body joint, and calories consumed during exercise are calculated in this process. We evaluate the performance of our method by benchmarking it to Fitbit, a smart watch well-known for health monitoring during exercise. The experimental results in this paper show that our method outperforms a state-of-the-art calorie assessment method, which we base on and improve, in terms of the error rate from Fitbit's ground-truth values.

## Introduction

It is suggested by several health experts that people should be concerned of their calorie intake and consumption (Hill et al. 2003). Nowadays, the assessment of calorie consumption remains challenging. There exists a gas analysis system for calorie consumption assessment (B Böhm, Hartmann, and H Böhm 2016), which seems highly accurate, but it needs large space and expensive devices. In addition, users of their system also lose freedom to move. Another method (Tsou and Wu 2015) was developed by Tsou and Wu where Kinect, a line of motion sensing input device that can detect the gesture of a whole body, is used for calorie assessment. This kind of device is expected to be extensively used in constructing rehabilitation applications in calorie assessment that are related to health promotion (Da Gama et al. 2015). In Tsou and Wu's method, the coordinates of body joints in 3D space are captured by Kinect and used to calculate the velocity of each joint movement, and then a kinetic energy for estimating calorie consumption. The method yields promising performance; however, there are still issues that can be improved, in particular, the issue that assessment does not take the body size of individual users into account.

In this paper, we propose an improved version of the method by Tsou and Wu. Note that in their method, kinetic energies are computed by using the velocities of body joints and the *standard* mass of each joint (a mass represents the portion of a joint of interest to the whole body, including muscles and bones attached to that joint). On the contrary, in our work, the mass of each body joint is derived by processing an image of the subject's body. In other words, calorie consumption assessment by our method takes the body size of each user into account. Following an existing protocol for system evaluation (Ryu, Kawahawa and Asami 2008), we use a reference device, Fitbit, to evaluate the assessment accuracy of our system.

## Existing Work

Nowadays, we could know how many calories a human consumes during walking by some smartphone applications. But accuracy is still in question. Most applications do not consider mass, which means they do not weight the importance of each body segment. Therefore, a method is required that adapts to the body size and weight of each individual user.

For the aforementioned exiting work on calorie consumption assessment based on gas analyzing, we stated that, based on their result, it is an accurate system. However, considering a high cost, largely needed space, it is impractical to adopt their approach to applications for promoting users' physical health through daily exercise or motion gaming.

Our work is mainly based on the aforementioned existing method by Tsou and Wu, in which Kinect is used to monitor users' activities and assess their calorie consumption. They showed error rates to a ground truth that is calorie consumption assessed by a reliable assessment tool, i.e., a heart rate monitor. In addition, the longer the training time, the less the error rate. They used kinetic energies of the body joints to build a regression function for estimating calorie consumption. The kinetic energy of each body joint is calculated as a multiplication of the joint's standard scale with the body weight. We conjecture that assessment can be improved if the body scale is measured specifically for each individual user.

## Methodology

According to Tsou and Wu's method, kinetic energy parameters are used to assess calorie consumption. This shows that such energies are related to the calorie consumption amount. Following their recipe, we also use kinetic energy parameters to assess calorie consumption.

The kinetic energy needs mass and velocity to calculate. In Tsou and Wu's method, the kinetic energy in each joint is used in multiple linear regressions for predicting calorie consumption. The assessment of mass, velocity, and calorie consumption are described in the subsections below, respectively.

### Mass

Tsou and Wu's method assumes that the shape of body is universal to all people while in our method, the system obtains mass by analyzing the body shape of each user specifically. Image processing is done on a depth image (An example is shown in Fig.1), where the ratio of each body segment to the whole body is computed and used to represent the mass percentage of each joint. By multiplying the mass percentage with the weight of the user, we obtain the mass of each part for calculation of the energy. To obtain the mass for each of Kinect's 20 joints, we used software called *ImageJ* to measure the ratio of the number of pixels in each joint's area to that in the whole body.

### Velocity

While a user is exercising, the system obtains his/her streaming skeleton data from Kinect (see *Figure 1*). The skeleton data represent 3D coordinates of all body joints in each row. We set the data frame rate to 25 fps. We derive the velocity of a given joint over a period of time by using the differentiation method.

The differentiation method is widely used in physics to obtain the average velocity over time. When the period is very short, we can regard this average velocity as the instantaneous velocity, i.e., the formula of which at time $t$ for joint $j$ is as follows:

$$v_{j,t} = \frac{ds}{dt} \tag{1}$$

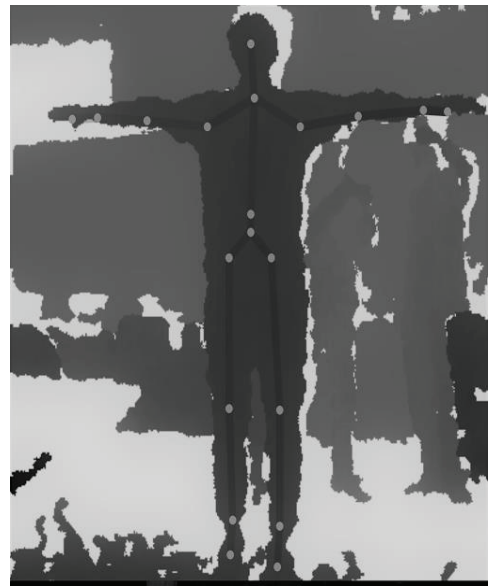

*Figure 1 An Example of a Depth Image*

where $ds$ is the distance that joint $j$ moves during the interval $[t - dt, t]$. For each joint, all instantaneous velocities are collected for the assessment of its kinetic energy.

**Calorie Consumption**

After obtaining the mass and velocity data of all body joints, we compute the kinetic energy for each one. As done in Tsou&Wu's original method, the values from three dimensions are used to calcuate the kinetic energy $E_j$ for joint $j$ as follows:

$$E_j = |E_{j,x} + E_{j,y} + E_{j,z}| \quad (2)$$

In Eq. (2), the parameters $E_{j,x}, E_{j,y}, E_{j,z}$ represent the kinetic energy in each dimension. Classical mechanics indicate that the kinetic energy $E$ of a particle of mass $M$ travelling at speed $V$ is given by $E = 1/2 MV^2$. As a result, Eq. (2) can be reformulated as follows:

$$E_j = \left|\frac{1}{2}M_j V_{j,x}^2 + \frac{1}{2}M_j V_{j,y}^2 + \frac{1}{2}M_j V_{j,z}^2\right| \quad (3)$$

In actual calculation, the velocity for each dimension in Eq. 3 is combined as one velocity. The parameter $M_j$ in this equation represents the mass of body joint $j$. This mass is obtained by Eq. 4 where $a_j$ is the ratio of joint $j$ in comparison to the area of the whole body, as described in Subsection **Mass**, and *weight* is the weight of a user of interest.

$$M_j = weight \times a_j \quad (4)$$

Note that $E_j$ in Eq. 2 is the kinetic energy at a given short period, e.g., 1 second. By accumulating this amount over the whole exercise session of, say, $T$ seconds, we obtain $K_j$ as an accumulated energy, or in other words the total energy spent for an exercise of interest (Eq. 5).

$$K_j = \sum_{t=1}^{T} E_{j,t}, \quad (5)$$

*Table 1 Twenty Joints in Kinect*

| Head | Center Shoulder | Left Shoulder | Right Shoulder |
|---|---|---|---|
| Left Elbow | Right Elbow | Left Wrist | Right Wrist |
| Left Hand | Right Hand | Spine | Center Hip |
| Left Hip | Right Hip | Left Knee | Right Knee |
| Left Ankle | Right Ankle | Left Foot | Right Foot |

where $E_{j,t}$ is

$$E_{j,t} = \frac{1}{2} M_j v_{j,t}^2 \quad (6)$$

Eq. 5 is applied to each of the 20 body parts (see *Figure 2*). Following the recipe in Tsou and Wu's method, calorie consumption (*CC*) is computed by using a multiple regression function having the resulting energies as input (Eq.7 where $b_0 \sim b_{20}$ indicate a bias and the coefficient for each dimension, respectively). The regression function is constructed in a training stage, in which *CC* from Fitbit is used as the dependent variable and the energies of all body parts are used as the independent variables in an analysis to find $b_0 \sim b_{20}$.

$$CC = b_0 + \sum_{j=1}^{20} b_j K_j \quad (7)$$

Eq. 7 is used for calculation of calorie consumption in our experiment for both Tsou and Wu's method and our method. For the former, $a_j$ in Eq. 4 is set to a standard scale of the human body. However, since in their work, some joints are combined, we need to separate them in order to have 20 joints as in our method. By checking the joints that didn't appear in their method, we found that the "body" part (30 percent of the whole body) mentioned in their method contains five parts of joint: Center Shoulder, Left and Right Shoulder, Spine, Center Hip and Left and Right Hip. As how each part contributes to Tsou and Wu's body remains unknown, we simply considered that all segments related to those joints share the same mass percentage, and when considering symmetric parts, we future divided the percentage into half. The 20 segments, each corresponding to one of Kinect's 20 joints (Table 1), and the standard scale for any subject are shown in Table 2.

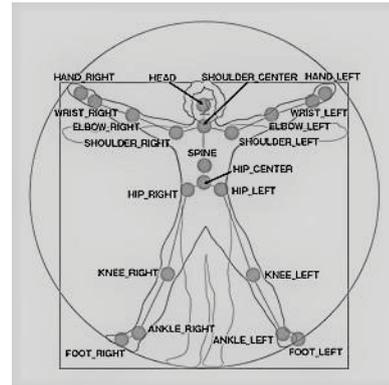

*Figure 2: The Concept Map of Joints in Kinect (NikkeiBP 2012)*

*Table 2 Standard Scale for Any Subject*

| Segment Name | Percentage | Accumulated %. | Joint Number |
|---|---|---|---|
| Head | 10% | 10% | 1 |
| Left, Right Elbow | 4%*2=8% | 18% | 2,3 |
| Left, Right Wrist | 3%*2=6% | 24% | 4,5 |
| Left, Right Hand | 2.5%*2=5% | 29% | 6,7 |
| Center Shoulder | 6% | 35% | 8 |
| Left, Right Shoulder | 3%*2=6% | 41% | 9,10 |
| Spine | 6% | 47% | 11 |
| Center Hip | 6% | 53% | 12 |
| Left, Right Hip | 3%*2=6% | 59% | 13,14 |
| Left, Right Knee | 10%*2=20% | 79% | 15,16 |
| Left, Right Ankle | 7%*2=14% | 93% | 17,18 |
| Left, Right Foot | 3.5%*2=7% | 100% | 19,20 |

## Experiment

We evaluated our system on two different sets of motions from broadcast gymnastics (BG): one by NHK (JPN[1]), Japan's national public broadcasting organization, and another by by Chinese Sports Government (CHN[2]). They are exercises that are popular and widely known among people in each respective nation. As a result, we used these sets of motions in our experiment.

## Process

For six subjects, each will be asked to do either JPN or CHN, depending on their choice, for construction of the prediction model. Figure 3 shows a subject doing an exercise in the experiment. This takes approximately 30 minutes. There are three steps as follows.

First, according to the method provided by Taylor (Taylor et al. 2012), before or after an experiment, a subject wears Fitbit and rests for 5 minutes. During such a period, the calorie consumption result from Fitbit is acquired. This data is required to ensure the measurement goes well by verifying whether the value in resting is not higher than the value in exercising.

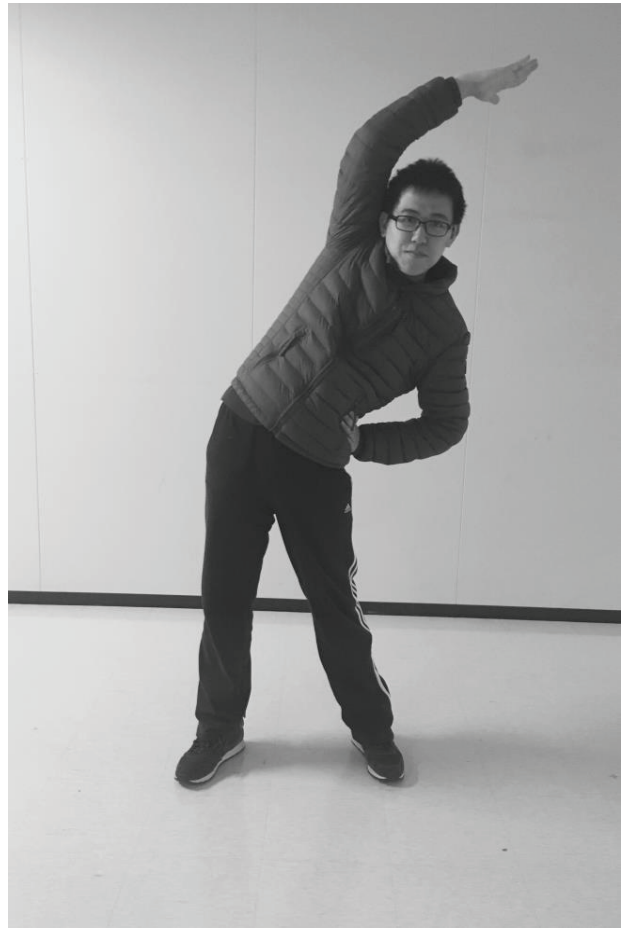

*Figure 3: A subject performing a broadcast gymnastics*

Second, the subject is asked to do JPN or CHN, either on their own after given a guidance or following an exercise video, while wearing Fitbit. Then after finishing exercise, the calorie consumption data from Fitbit are collected, and the first and second steps will be repeated twice.

Third, when the three cycles for the first and second steps are finished, a photo of the subject is taken with his/her hands stretched up. This photo contains 20 joints. It is used in image processing to obtain the mass scale of body joints.

---

[1] JPN Broadcast Gymnastics, 1st version, https://www.youtube.com/watch?v=b4SH_lap4ag

[2] CHN 9th National Broadcast Gymnastics official, http://www.iqiyi.com/w_19rqvi3qt9.html

### Data

There are two types of data in our experiment: the ground truth data from Fitbit and the mass data.

The ground truth is the calorie consumption assessed by Fitbit, both during the resting and exercising (engaging in JPN or CHN) time of the experiment. In this experiment, there are six subjects (three subjects from Japan, three subjects from China) for evaluating the prediction model of calorie consumption. Each subject did BG three times. The data are shown in Table 3, where $i$R represents the calorie loss the rest time before the $i^{st}$ exercise, and $i$E represents the calorie loss in the $i^{st}$ exercise. From this set of data, it can be seen that the calorie consumption in the rest situation (marked as 1R, 2R, 3R) is not more than the consumption in exercising (marked as 1E, 2E, 3E). The results from 1E, 2E, and 3E are used in comparison of the two prediction models.

*Table 3 Ground Truth Data from Fitbit(Data Unit: kcal)*

| Name | 1R | 1E | 2R | 2E | 3R | 3E |
|---|---|---|---|---|---|---|
| Sub.1 | 12 | 16 | 17 | 20 | 16 | 19 |
| Sub.2 | 9 | 25 | 9 | 26 | 7 | 29 |
| Sub.3 | 6 | 14 | 7 | 19 | 6 | 15 |
| Sub.4 | 7 | 20 | 9 | 17 | 8 | 19 |
| Sub.5 | 9 | 22 | 14 | 26 | 6 | 21 |
| Sub.6 | 11 | 31 | 13 | 30 | 26 | 33 |

*Table 4 Example of Mass Data of a Subject*

| Segment Name | Percentage | Total |
|---|---|---|
| Head | 5.76% | 5.76% |
| Center Shoulder | 9.99% | 15.75% |
| Left Shoulder | 5.49% | 21.24% |
| Right Shoulder | 5.49% | 26.73% |
| Left Elbow | 3.07% | 29.80% |
| Right Elbow | 3.07% | 32.87% |
| Left Wrist | 1.40% | 34.27% |
| Right Wrist | 1.40% | 35.67% |
| Left Hand | 1.05% | 36.72% |
| Right Hand | 1.05% | 37.77% |
| Spine | 10.47% | 48.24% |
| Center Hip | 3.64% | 51.88% |
| Left Hip | 4.36% | 56.24% |
| Right Hip | 4.36% | 60.60% |
| Left Knee | 9.64% | 70.24% |
| Right Knee | 9.64% | 79.88% |
| Left Ankle | 7.10% | 86.98% |
| Right Ankle | 7.10% | 94.08% |
| Left Foot | 2.96% | 97.04% |
| Right Foot | 2.96% | 100.00% |

The mass data shows the body scale of each subject. It is unique to each subject as shown for example in Table 4. This data is multiplied by the subject's weight for each segment (Eq. 4) in order to obtain the segment's mass.

**Performance Metric:**

The metric for performance evaluation is shown in Eq. 8. This metric shows the error rate in calorie consumption assessment for the $n^{th}$ subject in the $i^{th}$ exercise where $E_{fn}^i$ is the result from Fitbit and $E_{kn}^i$ is the result from a prediction model of interest.

$$Error\_rate(n,i) = \frac{|E_{fn}^i - E_{kn}^i|}{E_{fn}^i} \quad (8)$$

### Results and Analysis

The results are shown in three parts: error results, cross validation results, and statistical test results. Error results are the evaluation results over training data, which indicate that our method outperforms Tsou and Wu's method. Crossvalidation results ensure that the proposed should work well even on unseen data. Statistical test results indicate that there is a statistically significant difference between the two methods in cross validation.

### Error Results

We compared CC measured by Tsou and Wu's and by our method to the ground truth provided by Fitbit. We benchmarked the two methods using the error rate (Eq. 8). Table 5 shows that our method yields less error rate than Tsou and Wu's method. In addition, the error rate of our method is obviously smaller than Tsou and Wu's method, which means we have successfully improved state-of the art Tsou and Wu's method in predicting calorie consumption.

*Table 5 Error Rates of Our Method and*
*Tsou & Wu's Method over the Training Data*

| Subject | Ours | Tsou & Wu's |
|---|---|---|
| 1 | $1.39*10^{-5}$ | $1.79*10^{-5}$ |
| 2 | $4.35*10^{-5}$ | $1.93*10^{-2}$ |
| 3 | $1.86*10^{-5}$ | $3.23*10^{-5}$ |
| 4 | $1.54*10^{-5}$ | $3.58*10^{-5}$ |
| 5 | $1.54*10^{-5}$ | $2.14*10^{-5}$ |
| 6 | $3.18*10^{-5}$ | $4.32*10^{-5}$ |

## Cross Validation Results

In order to confirm the accuracy of the prediction model, we ran a 3-fold cross validation. In this cross validation, part of the data of all subjects (e.g., data from the first and second exercises for all subjects) are used for constructing a prediction model for each method, then the prediction models are tested on the remaining data (e.g., referring to the example above, data from the third exercise); this is done three times, each with a different combination of training and testing data, for each method in order to obtain the average result.

Table 6 shows the error rates in cross validation. Note that the error rate of our method (Tsou and Wu's) at the $i^{st}$ exercise, Ours$_i$ (Tsou & Wu$_i$), shows the performance of the prediction model based on the corresponding method using the data in the remaining exercises for training. As can be seen from the table, the error rates of the proposed method are in most cases less than Tsou and Wu's method.

## Statistical Test Results

We conducted a Wilcoxon signed-rank test to find whether there is a statistically significant difference between error rates from the two methods. The resulting $p$ value is 0.00854, which is less than 0.01, indicating that there is a significant difference at the confident interval of 99%. As a result, we can state that our attempt to improve Tsou and Wu's method through personalization of the user's mass scale is successful.

## Conclusions and Future Work

We have presented a personalized method for calorie consumption assessment using Kinect based on the unique shape of each user. Kinect can produce skeleton data for analyzing the movements of body joints that lead to the velocity of each joint, and depth images that lead to mass data that are unique to each subject, both of which enable kinetic energy calculation. We build a prediction model based on the results from the ground truth data that connects the kinetic energies from Kinect. By comparing to the prediction model by Tsou and Wu, which uses standard scale mass data on every subject, our method utilizing personalized mass data outperforms Tsou and Wu's method, both in evaluation over training data and in evaluation using cross validation.

In future work, we will employ this method to monitoring the health state of motion-game players. This can be done by constructing a calorie consumption system that uses Kinect and a ground truth device for a prediction model, and by considering the effect of the amount of exercises (Slentz et al. 2004). We will also add a potential energy into the assessment formulas and estimate post-exercise calorie burn. In addition, our method can be used for health monitoring during full-body motion gaming to promote a healthy exercise while preventing injuries.

## Acknowledgement

The authors wish to thank members of Intelligent Computer Entertainment Lab, Ritsumeikan University, for their helps in the experiment and suggestions.

*Table 6 Error Rates of Our Method and Tsou & Wu's Method for Each Testing Data in a Three-Fold Cross Validation*

| Subject | Ours$_1$ | Ours$_2$ | Ours$_3$ | Tsou & Wu's$_1$ | Tsou & Wu's$_2$ | Tsou & Wu's$_3$ |
|---|---|---|---|---|---|---|
| 1 | 0.2106 | **0.2196** | 0.1408 | **0.1537** | 0.1464 | **0.1266** |
| 2 | **0.0303** | **0.0208** | 0.1429 | 0.3009 | 0.5538 | 0.3447 |
| 3 | **0.1166** | **0.0834** | 0.3677 | 0.3409 | 0.5433 | **0.4870** |
| 4 | 0.3960 | **0.1634** | **0.0501** | **0.2506** | 0.3803 | 0.3096 |
| 5 | **0.1045** | **0.0458** | 0.1501 | 0.2227 | 0.2724 | 0.1927 |
| 6 | **0.0529** | 0.6126 | 0.3941 | 0.2040 | **0.5131** | **0.4292** |
| Avg. | | 0.1835 | | | 0.3207 | |